\documentclass[conference]{IEEEtran}
\IEEEoverridecommandlockouts
\usepackage{cite}
\ifCLASSINFOpdf
\else
  \usepackage[dvips]{graphicx}
  \graphicspath{{../eps/}}
  \DeclareGraphicsExtensions{.eps}
\fi
\usepackage{amsmath}
\usepackage{bm}
\usepackage{amssymb}
\usepackage{array}
\usepackage{multirow}
\usepackage{url}
\usepackage{algorithmic}
\usepackage{algorithm}
\usepackage{diagbox}
\usepackage{subfigure}
\usepackage{etoolbox}
\usepackage{amsmath}

\makeatletter
\patchcmd{\@makecaption}
  {\scshape}
  {}
  {}
  {}
\makeatother
\usepackage{fancyhdr}

\begin{document}
%
\title{DFT-s-OFDM with Chirp Modulation}



%
\author{\IEEEauthorblockN{Yujie Liu\IEEEauthorrefmark{1},
Yong Liang Guan\IEEEauthorrefmark{1},
David Gonz\'alez G.\IEEEauthorrefmark{2},
Halim~Yanikomeroglu\IEEEauthorrefmark{3}}
\IEEEauthorblockA{\IEEEauthorrefmark{1}Continental-NTU Corporate Lab,
Nanyang Technological University, Singapore}
\IEEEauthorblockA{\IEEEauthorrefmark{2}Wireless Communications Technologies Group, Continental AG, Germany}
\IEEEauthorblockA{\IEEEauthorrefmark{3}Department of Systems and Computer Engineering, Carleton University, Ottawa, Canada\\
Email: \{yujie.liu, eylguan\}@ntu.edu.sg, david.gonzalez.g@ieee.org, halim@sce.carleton.ca}


\thanks{This research/project is supported by A*STAR under the RIE2025 Industry Alignment Fund - Industry Collaboration Projects (IAF-ICP) Funding Initiative (Award: I2501E0045), as well as cash and in-kind contribution from the industry partner(s).}}


\maketitle
\thispagestyle{fancy}
\lfoot{This paper has been accepted by IEEE PIMRC 2025. Copyright (c) 20xx IEEE. Personal use of this material is permitted. However, permission to use this material for any other purposes must be obtained from the IEEE by sending a request to pubs-permissions@ieee.org}

\begin{abstract}
In this paper, a new waveform called discrete Fourier transform spread orthogonal frequency division multiplexing with chirp modulation (DFT-s-OFDM-CM) is proposed for the next generation of wireless communications. The information bits are conveyed by not only $Q$-ary constellation symbols but also the starting frequency of chirp signal. It could maintain the benefits provided by the chirped discrete Fourier transform spread orthogonal frequency division multiplexing \mbox{(DFT-s-OFDM)}, \emph{e.g.}, low peak-to-average power ratio (PAPR), full frequency diversity exploitation, etc. Simulation results confirm that the proposed DFT-s-OFDM-CM could achieve higher spectral efficiency while keeping the similar bit error rate (BER) to that of chirped DFT-s-OFDM. In addition, when maintaining the same spectral efficiency, the proposed DFT-s-OFDM-CM with the splitting of information bits into two streams enables the use of lower-order constellation modulation and offers greater resilience to noise, resulting in a lower BER than the chirped DFT-s-OFDM.
\end{abstract}


\section{Introduction}
Wireless communications, as one of the most remarkable technological advancements in modern society over the past decades, has a revolutionary impact on people's living, studying, working, socializing, entertaining, etc. Since $1980$s, five generations of mobile communications have been standardized, with each generation emerging a decade approximately. The sixth generation (6G) of mobile communications is likely to be commercially deployed by $2030$, offering ultra-high data rates, ultra-low latency, ultra-high reliability, ultra-high spectral efficiency, etc. Compared to the current fifth generation (5G) of communications, 6G will introduce enhanced capabilities such as integrated sensing and communications (ISAC) \cite{9468975} and support more demanding applications like high-mobility communications \cite{afdm_twc,10897935}.

Transmission waveform is an important component of any modern wireless communication/sensing networks. Specifically, orthogonal frequency division multiplexing (OFDM)~\cite{9468975} offers several outstanding advantages, \emph{e.g.}, its robustness to frequency selective fading, simple implementation using fast Fourier transform (FFT), high-data rate transmission, etc. It is thus widely adopted in wireless communication systems for downlink communications, including the Wireless Fidelity~(WiFi) networks, Long Term Evolution (LTE) and New Radio (NR) in the fourth and fifth generations of communications. In contrast, discrete Fourier transform spread OFDM (DFT-s-OFDM) \cite{7744818} is chosen for uplink communications due to its low peak-to-average-power ratio (PAPR). However, the communication performance of OFDM and DFT-s-OFDM tends to deteriorate in high-mobility scenarios. Chirp signal, also known as (\emph{a.k.a.}) frequency modulated continuous wave (FMCW)~\cite{8943249}, is broadly used in radar systems to measure the distance, velocity, and angle of targets. One of the popular internet-of-things (IoT) protocols, called long-range radio \mbox{(LoRa)}~\cite{10323409,8883217}, uses chirp spread spectrum (CSS) modulation to enable low-power and long-range communications. However, LoRa exhibits low spectral efficiency, which makes it unsuitable for applications that demand high data rates.

Recently, several new waveforms have been proposed in the literature that integrate communication waveforms (\emph{e.g.}, \mbox{OFDM}, DFT-s-OFDM) with sensing waveforms (\emph{e.g.}, chirp signals), including affine frequency division multiplexing (AFDM) \cite{afdm_twc} and chirped DFT-s-OFDM \cite{10897935}. Both AFDM and chirped DFT-s-OFDM can leverage full frequency diversity and achieve performance comparable to that of orthogonal time frequency space (OTFS) \cite{afdm_twc,10897935}. Furthermore, compared to AFDM and OTFS, chirped DFT-s-OFDM offers an additional advantage - lower PAPR, making it well-suited for low-power devices. To further enhance spectral efficiency, index modulation (IM) has been introduced to AFDM, including AFDM with IM (AFDM-IM) in \cite{10845819,10342712,10570960}, AFDM with pre-chirp-domain IM (AFDM-PIM) \cite{10975107}, AFDM with chirp-permutation IM (AFDM-CPIM) \cite{10943004}, etc. Regarding AFDM-IM, the information bits are encoded into not only $Q$-ary constellation symbols but also the activation states of chirp subcarriers in the discrete affine Fourier (DAF) domain. The degree of freedom in selecting the second chirp parameter $c_2$ in AFDM allows information bits to be embedded in the second chirp signal, leading to the development of AFDM-PIM \cite{10975107} and AFDM-CPIM \cite{10943004}. However, the variants of AFDM still suffer high PAPR. Though the PAPR of AFDM-PIM could be reduced in \cite{10812762}, it is still higher than that of $Q$-ary constellation symbols. Therefore, to the best of the authors' knowledge, there is currently no modulation waveform in the literature simultaneously offering high spectral efficiency, superior performance, and low PAPR.


\begin{figure*}[htbp]
	\centering
	\includegraphics[width=12cm]{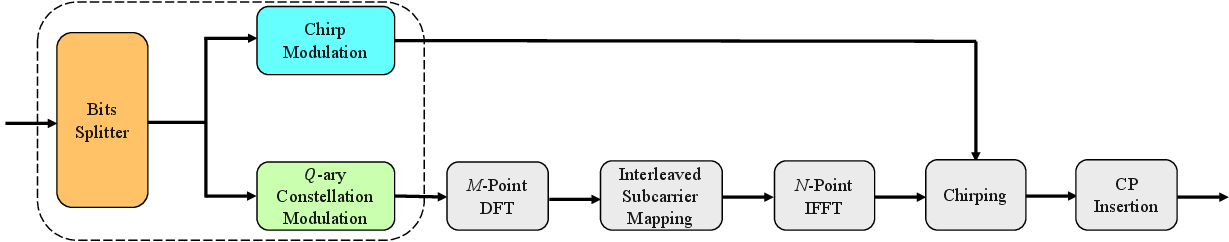}
	\caption{Block diagram of proposed DFT-s-OFDM-CM for user $u$.}
	\label{Fig.dft_chirp_s_ofdm}
\end{figure*}

\begin{figure*}[htbp]
	\centering
	\includegraphics[width=12cm]{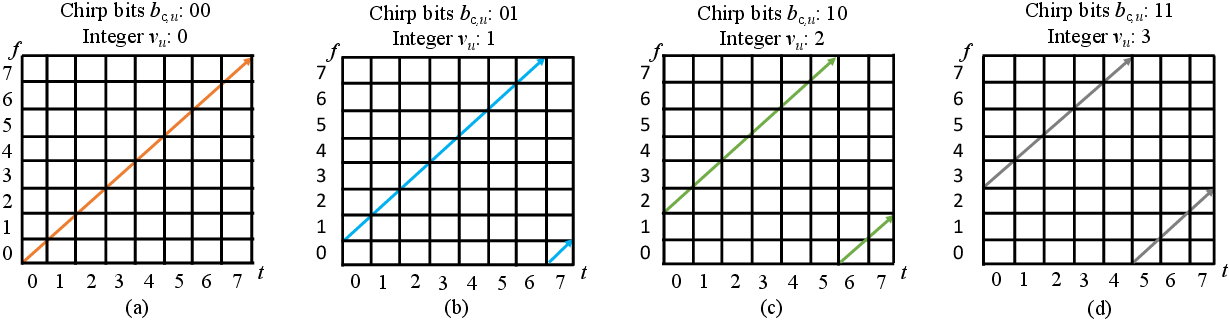}
	\caption{Illustrative example: chirp modulation. Information bits are encoded into the starting frequency of chirp signal.}
	\label{Fig.cm}
\end{figure*}

In this paper, to enhance the spectral efficiency of DFT-s-OFDM, chirp modulation is integrated with DFT-s-OFDM, leading to a new transmission waveform called DFT-s-OFDM with chirp modulation (DFT-s-OFDM-CM). The information bits are embedded on not only the $Q$-ary constellation symbols, but also the starting frequency of chirp signal. The contributions of this paper are outlined in the following:
\begin{itemize}
\item To the best of the authors' knowledge, this is the first transmission waveform in the literature to introduce the concept of CSS modulation to DFT-s-OFDM, offering both high spectral efficiency and low PAPR. In contrast, the existing waveforms including AFDM-IM~\cite{10845819,10342712,10570960}, AFDM-PIM~\cite{10975107}, and AFDM-CPIM \cite{10943004} could achieve high spectral efficiency but suffer higher PAPR. The existing waveforms such as chirped DFT-s-OFDM \cite{10897935} offers lower PAPR but their chirp signal does not convey information bits, leading to lower spectral efficiency.
\item By properly designing chirp modulation with DFT-s-OFDM, additional information bits can be encoded into the starting frequency of chirp signal without sacrificing the performance. Simulation results show that the proposed DFT-s-OFDM-CM could increase the spectral efficiency by $50\%$ while maintaining the similar bit error rate (BER) to that of chirped DFT-s-OFDM \cite{10897935}.
\item Thanks to the splitting of information bits into two separate streams, lower-order constellation modulation could be enabled for DFT-s-OFDM-CM, resulting in higher resilience to noise than the chirped DFT-s-OFDM~\cite{10897935}. Simulation results show that when maintaining the same spectral efficiency, the proposed DFT-s-OFDM-CM presents lower BER than that of chirped DFT-s-OFDM~\cite{10897935}, with an approximately $2\,\rm{dB}$ gain.
    \item The pairwise error probability (PEP) analysis is conducted for the proposed DFT-s-OFDM-CM and the derived BER upper bound is found to approach the simulated BER especially at medium to high $E_{\mathrm{b}}/N_0$. Similarly to AFDM \cite{afdm_twc} and chirped DFT-s-OFDM \cite{10897935}, diversity order analysis indicates that the proposed DFT-s-OFDM-CM is able to achieve full frequency diversity.
\end{itemize}

The structure of this paper is as follows. Section II provides a description of proposed DFT-s-OFDM-CM along with chirp modulation optimization. Section III conducts performance analysis for it using the PEP analysis of maximum likelihood (ML) equalizer. Simulation results are provided in Section~IV, including an extension of this concept to AFDM. This paper is concluded in Section V.

\section{DFT-s-OFDM with Chirp Modulation}
\subsection{System Model}
Considering uplink multiple access communications with $U$ users, the block diagram of proposed DFT-s-OFDM-CM of user $u$ is illustrated in Fig. \ref{Fig.dft_chirp_s_ofdm}. Assume the number of total bits for user $u$ is $b_u$, which is then divided into two parts, $b_{\mathrm{c},u}$ bits for chirp modulation and $b_{\mathrm{s},u}$ bits for $Q$-ary constellation modulation, respectively. Similarly to CSS modulation \cite{10323409}, $b_{\mathrm{c},u}$ bits are converted to an integer value denoted as $\nu_u$. Define the chirp modulation order as $P$ and the number of bits $b_{\mathrm{c},u}$ is determined by $\log_2P$.

The linear up-chirp signal is written as
\begin{equation}
c[n]=e^{j\pi c_{\mathrm{r}}n^2},
\label{c_u_n}
\end{equation}
where $c_{\rm{r}}$ is the chirp rate and $n=0,1,\cdots,N-1$ with $N$ being the number of total subcarriers. In this paper, $c_{\rm{r}}=1/N$ is considered \cite{10897935}. Note that frequency of a signal is the derivative of its phase with respect to time. Accordingly, the frequency of chirp signal \eqref{c_u_n} with $c_{\rm{r}}=1/N$ at time instant $n$ corresponds to frequency bin $n$. The chirp signal $c[n]$ is illustrated in Fig. \ref{Fig.cm}a using an orange line with an arrow, indicating that it begins at a frequency of zero and increases linearly over time with a chirp rate of $1/N$ ($N=8$). The chirp bits are encoded into the starting frequency of chirp signal by circularly shifting the elements in array $c[n]$. The modulated chirp signal of user $u$ can be expressed as
\begin{equation}
c_u[n]=\mathrm{circshift}\{c[n],-\nu_u\},
\label{c_u_n_circ}
\end{equation}
where $\mathrm{circshift}$ denotes Matlab function to circularly shift the elements in array $c[n]$ by $-\nu_u$ positions. Fig. \ref{Fig.cm} illustrates chirp modulation for $P=4$. When two bits $b_{\mathrm{c},u}$ are $00$, corresponding to $\nu_u =0$, the modulated chirp signal remains identical to the original chirp signal $c[n]$ as depicted in Fig.~\ref{Fig.cm}a. For bits $b_{\mathrm{c},u}=01$, $10$, and $11$, the original chirp signal $c[n]$ is circularly left-shifted by $\nu_u =1$, $\nu_u =2$, and $\nu_u =3$ positions, respectively. This results in modulated chirp signals whose starting frequencies correspond to frequency bin $\nu_u=1$, $\nu_u=2$, and $\nu_u=3$, as illustrated in Fig. \ref{Fig.cm}b, Fig.~\ref{Fig.cm}c, and Fig.~\ref{Fig.cm}d, respectively. Note that unlike chirp modulation in Fig.~\ref{Fig.cm}, index modulation encodes information by selecting active indices (\emph{e.g.}, subcarriers or antennas) without directly altering the signal's amplitude, frequency, or phase.


Following $Q$-ary constellation modulation, $b_{\mathrm{s},u}$ bits are mapped to a data symbol vector of length $M$, \emph{i.e.,} $\mathbf{x}_u=[x_u[0],x_u[1],\cdots,x_u[M-1]]^T$. The DFT-s-OFDM signal of user $u$ can be written as
\begin{equation}
\mathbf{s}_u=\mathbf{F}_N^H\mathbf{P}_u\mathbf{F}_M\mathbf{x}_u,
\end{equation}
where $\mathbf{F}_M$, $\mathbf{F}_N$, and $\mathbf{P}_u$ are $M$-point discrete Fourier transform (DFT) matrix, $N$-point FFT matrix, and interleaved subcarrier mapping matrix of user $u$ \cite{10897935}. $\mathbf{P}_u=\mathbf{I}_N(:,I_u:\frac{N}{M}:N)$ is defined, with $\mathbf{I}_N$ being an $N\times N$ identity matrix, $I_u$ ($1\leq I_u\leq N/M$) the subcarrier index of user $u$, and $N/M$ an integer. By incorporating the modulated chirp signal, the resulted DFT-s-OFDM-CM signal is formulated as
\begin{equation}
   \mathbf{s}_{\mathrm{c},u}=\mathbf{C}_u\mathbf{F}_N^H\mathbf{P}_u\mathbf{F}_M\mathbf{x}_u,
\end{equation}
with $\mathbf{C}_u=\mathrm{diag}\{\mathbf{c}_u\}$ and $\mathbf{c}_u=[c_u[0],c_u[1],\cdots,c_u[N-1]]$.
A cyclic prefix (CP) is then added to $ \mathbf{s}_{\mathrm{c},u}$ before transmission. The proposed DFT-s-OFDM-CM has a modulation complexity of $M\log_2M+N\log_2N+N$.

The spectral efficiency of (chirped) DFT-s-OFDM and proposed DFT-s-OFDM-CM can be given by
\begin{equation}
\mathrm{SE}_{\mathrm{DFT-s-OFDM}}=\frac{UM\log_2 Q}{N},
\label{se_dft_s_ofdm}
\end{equation}
\begin{equation}
\mathrm{SE}_{\mathrm{DFT-s-OFDM-CM}}=\frac{UM\log_2 Q+U\log_2 P}{N}.
\label{se_dft_s_ofdm_cm}
\end{equation}
According to \eqref{se_dft_s_ofdm} and \eqref{se_dft_s_ofdm_cm}, the proposed DFT-s-OFDM-CM could transmit additional bits of $U\log_2 P$ for chirp modulation, contributing to higher spectral efficiency.

Considering the delay-Doppler channel in \cite{afdm_twc} and \cite{10897935} with $L$ paths, the channel gain of $l$-th ($l=1,2,\cdots,L$) path at $n$-th time instant for user $u$ can be expressed as:
\begin{equation}
h_u[n,l]=\sum\nolimits_{p=1}^{L}h_{u,p}e^{j2\pi v_{u,p}n/N}\delta(l-l_{u,p}),
\label{h_n_l}
\end{equation}
where $h_{u,p}$, $v_{u,p}$, and $l_{u,p}$ are the channel amplitude, normalized Doppler shift, and integer delay of $p$-th path for user $u$, respectively \cite{10897935}. Note that $h_{u,p}$ is characterized as an \emph{i.i.d.} complex Gaussian random variable with zero mean and variance of $1/L$. $v_{u,p}$ is randomly chosen within the range of $-\bar{f}_{\textrm{max}}$ and $\bar{f}_{\textrm{max}}$, with $\bar{f}_{\textrm{max}}$ denoted as the normalized maximum Doppler frequency \cite{10897935}. The different channel paths are assumed to have different integer delays. Nevertheless, like in \cite{afdm_twc} and \cite{10897935}, the extension of \eqref{h_n_l} to the scenarios with  multiple paths sharing the identical integer delay is feasible. The time-domain channel matrix is given by
\begin{equation}
\mathbf{H}_{\rm{t},u}=\sum\nolimits_{l=1}^{L}h_{u,l}\mathbf{D}_{u,l}\bm{\Pi}^{l},
\end{equation}
where $\mathbf{D}_{u,l}=\mathrm{diag}\{1,e^{j2\pi v_{u,l}/N},\cdots,e^{j2\pi v_{u,l}(N-1)/N}\}$ and the forward cyclic-shift matrix $\bm{\Pi}^{l}$ is described in eq. (25) of~\cite{afdm_twc}.
After the removal of CP, considering $U$ users, the received time-domain DFT-s-OFDM-CM signal is given by
\begin{align}
\mathbf{r}=\sum\nolimits_{u=1}^{U}\mathbf{H}_{\mathrm{t},u}\mathbf{C}_u\mathbf{F}_N^H\mathbf{P}_u\mathbf{F}_M\mathbf{x}_u+\mathbf{w},
\label{r}
\end{align}
where $\textbf{w}$ is modeled as an additive white Gaussian noise vector of variance $\sigma^2$.

 \begin{figure}[htbp]
	\centering
	\includegraphics[width=8cm]{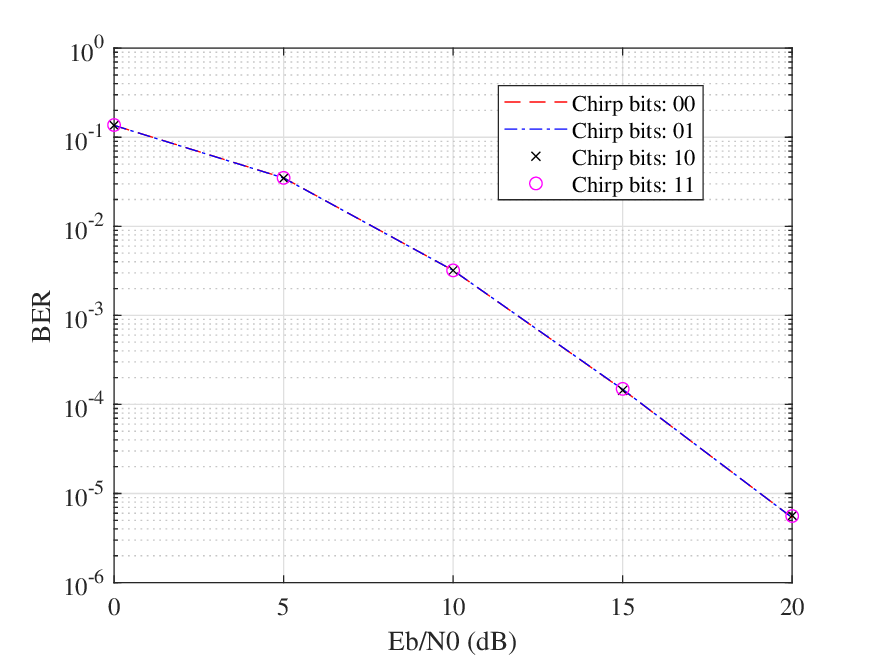}
	\caption{BERs of DFT-s-OFDM-CM with chirp bits shown in Fig. \ref{Fig.cm}, with $N=8$, $M=2$, $U=4$, $Q=2$, $L=3$, and $v=500\,\rm{km/h}$. DFT-s-OFDM-CM with chirp bits $00$ turns out to be the chirped DFT-s-OFDM~\cite{10897935}.}
\label{ber_diff_chirp}
\end{figure}

\begin{table}[htbp]
	\centering
	\caption{PAPRs of DFT-s-OFDM-CM with chirp bits shown in Fig. \ref{Fig.cm} calculated using $N=8$, $M=2$, and $U=4$.}
	\begin{tabular}{|c|c|c|c|c|}
		\hline
Chirp bits $b_{\mathrm{c},u}$ & $00$ & $01$ & $10$ & $11$ \\\hline
$Q$-ary PSK & \multicolumn{ 4}{c|}{$0\,\rm{dB}$}  \\\hline
	\end{tabular}
\label{tab1}
\end{table}

 Fig.~\ref{ber_diff_chirp} shows the BERs of DFT-s-OFDM-CM with four different chirp bits shown in Fig. \ref{Fig.cm}. Their BERs are very close to each other, implying that the starting frequency of chirp signal would not affect the BER. Note that DFT-s-OFDM-CM with chirp bits $00$ turns out to be the chirped DFT-s-OFDM in \cite{10897935}. In addition, Table~\ref{tab1} confirms that the chirp signal with different starting frequencies would not change the PAPR of DFT-s-OFDM-CM. As a result, it is feasible to encode information bits on the starting frequency of chirp signal, while without affecting the performance and PAPR. Similar to (chirped) DFT-s-OFDM \cite{10897935}, the PAPR of DFT-s-OFDM-CM with interleaved subcarrier mapping corresponds to that of $Q$-ary constellation symbols. When $Q$-ary phase shift keying (PSK) modulation is considered, DFT-s-OFDM-CM could attain a PAPR of $0\,\rm{dB}$. Besides, Fig. 4a in \cite{10897935} shows that (chirped) DFT-s-OFDM exhibits lower PAPR than OTFS, OFDM, and AFDM; thus, the proposed DFT-s-OFDM-CM, as its variant, is also expected to yield lower PAPR.




\subsection{Chirp Modulation Optimization}
With an improper chirp modulation order $P$, there may be multiple solutions (multiple pairs of $\widehat{\mathbf{c}}$ and $\widehat{\boldsymbol{x}}$) to satisfy the minimization problem below, leading to ambiguities:
\begin{equation}
\left(\widehat{\mathbf{c}}, \widehat{\boldsymbol{x}}\right)=\min _{\tilde{\mathbf{c}}_{u}\in \mathbb{P}, \widetilde{\boldsymbol{x}}_u \in \mathbb{Q}}\left\|\sum_{u=1}^U\left(\boldsymbol{s}_{\mathrm{c},u}-\tilde{\mathbf{C}}_u \mathbf{F}_N^H \mathbf{P}_u \mathbf{F}_M \widetilde{\boldsymbol{x}}_u\right)\right\|,
\label{ml_c_x}
\end{equation}
with $\widehat{\mathbf{c}}=[\widehat{\mathbf{c}}_1^T,\widehat{\mathbf{c}}_2^T,\cdots,\widehat{\mathbf{c}}_{U}^T]^T$ and $\widehat{\boldsymbol{x}}=[\widehat{\boldsymbol{x}}_1^T,\widehat{\boldsymbol{x}}_2^T,\cdots,\widehat{\boldsymbol{x}}_U^T]^T$.
Hence, the chirp modulation order $P$ needs to be optimized to result in a unique solution to \eqref{ml_c_x}. To simply the minimization problem \eqref{ml_c_x}, multiple users are assumed to have the same chirp signal $\widehat{\mathbf{c}}_0$, yielding the new optimization problem
\begin{equation}
\left(\widehat{\mathbf{c}}_0, \widehat{\boldsymbol{x}}\right)=\min _{\tilde{\mathbf{c}}_{0} \in \mathbb{P}, \widetilde{\boldsymbol{x}}_u \in \mathbb{Q}}\left\|\sum_{u=1}^U\left(\boldsymbol{s}_{\mathrm{c},u}-\tilde{\mathbf{C}}_0 \mathbf{F}_N^H \mathbf{P}_u \mathbf{F}_M \widetilde{\boldsymbol{x}}_u\right)\right\|
\label{ml_c_x_same}
\end{equation}
with $\tilde{\mathbf{C}}_0=\mathrm{diag}\{\tilde{\mathbf{c}}_0\}$.
Define the candidate of chirp modulation order as $\tilde{P}$. \eqref{ml_c_x_same} is first performed with an initial value of $\tilde{P}=N$. If there are multiple solutions, \eqref{ml_c_x_same} is re-performed with $\tilde{P}$ reduced by half next round, until a unique solution is found. The final value of $\tilde{P}$ is denoted as $P^{\star}$. Then, by substituting $P^{\star}$ to \eqref{ml_c_x}, it can be found that $P^{\star}$ could lead to a unique solution to \eqref{ml_c_x} as well. To prevent ambiguity, the value of $P$ should be selected such that $P\leq P^{\star}$, as ambiguities would occur when $P>P^{\star}$.

Note that $c[n]$ considered in this paper is an up-chirp signal with frequency increasing over time. Nevertheless, down-chirp signal with frequency decreasing over time can also be considered. The modulated down-chirp signal is written as
\begin{equation}
c_{\rm{d},u}[n]=\mathrm{circshift}\{c_{\rm{d}}[n],-\nu_u\},
\label{c_u_d}
\end{equation}
with $c_{\rm{d}}[n]=e^{-j\pi c_{\mathrm{r}}n^2}$. To further improve spectral efficiency, chirp modulation can incorporate both up-chirp and down-chirp signals, with an additional bit per user encoded to indicate the index of either an up-chirp or down-chirp.

\section{Performance Analysis}
According to \cite{10050811}, the single-user BER upper bound can be used as the benchmark of multi-user BER performance. The single-user BER upper bound of DFT-s-OFDM-CM is thus derived by using PEP analysis of ML equalizer below.

\begin{figure}[htbp]
	\centering
	\includegraphics[width=8cm]{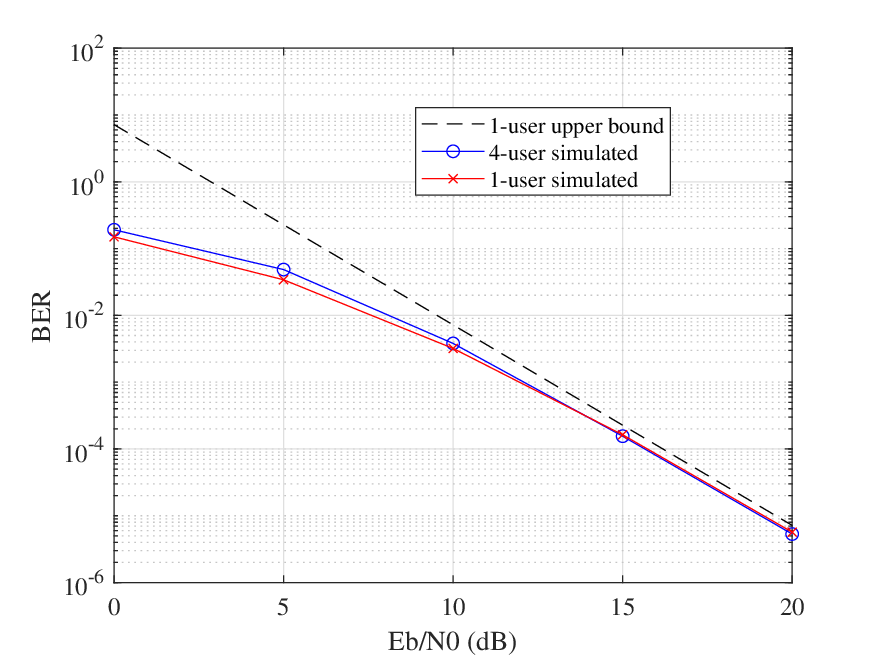}
	\caption{Derived BER upper bound and simulated BERs of proposed DFT-s-OFDM-CM, with $N=8$, $M=2$, $Q=2$, and $P=2$.}
\label{ber_ub}
\end{figure}

Considering a single-user scenario, the received DFT-s-OFDM-CM signal \eqref{r} can be rewritten as
\begin{align}
\mathbf{r}&=\sum\nolimits_{l=1}^{L}h_l\mathbf{D}_l\bm{\Pi}^{l}\mathbf{C}\mathbf{F}_N^H\mathbf{P}\mathbf{F}_M\mathbf{x}+\mathbf{w},\notag\nonumber\\
&=\sum\nolimits_{l=1}^{L}\mathbf{E}_l(\mathbf{c},\mathbf{x})h_l+\mathbf{w}=\mathbf{E}(\mathbf{c},\mathbf{x})\mathbf{h}+\mathbf{w}
\end{align}
where $\mathbf{E}=[\mathbf{E}_1,\mathbf{E}_2,\cdots,\mathbf{E}_L]$ and $\mathbf{h}=[h_1,h_2,\cdots,h_L]^T$, with $\mathbf{E}_l=\mathbf{D}_l\bm{\Pi}^{l}\mathbf{C}\mathbf{F}_N^H\mathbf{P}\mathbf{F}_M\mathbf{x}$. The ML equalizer is used to estimate chirp signal $\mathbf{c}$ and constellation symbol vector $\mathbf{x}$:
\begin{equation}
\left( \hat{\mathbf{c}},\hat{\mathbf{x}}\right)=\min_{\tilde{\mathbf{c}} \in \mathbb{P}, \widetilde{\mathbf{x}} \in \mathbb{Q}}\lVert \mathbf{r}-\mathbf{E}(\tilde{\mathbf{c}},\tilde{\mathbf{x}})\mathbf{h} \rVert.
\end{equation}
Denote $\mathbf{a}=[\mathbf{c},\mathbf{x}]^T$. The pairwise error event is defined as $\{\mathbf{a}\rightarrow\hat{\mathbf{a}}\}$, where $\mathbf{a}$ are the true values and $\hat{\mathbf{a}}$ are erroneously estimated values using ML equalizer. Define $\bm{\Theta}(\mathbf{a},\hat{\mathbf{a}})=(\mathbf{E}(\mathbf{a})-\mathbf{E}(\hat{\mathbf{a}}))^H(\mathbf{E}(\mathbf{a})-\mathbf{E}(\hat{\mathbf{a}}))$ and $\gamma=\frac{1}{\sigma^2}$. Denote the rank of $\bm{\Theta}(\mathbf{a},\hat{\mathbf{a}})$ and its non-zero eigenvalues as $R$ and $\{\lambda_1,\lambda_2,\cdots,\lambda_R\}$. Following the similar derivations in \cite{10897935} and \cite{10050811}, the BER upper bound is formulated as
\begin{equation}
P_e\leq \frac{1}{f}\sum\nolimits_{\mathbf{a}}\sum\nolimits_{\hat{\mathbf{a}},\mathbf{a}\neq\hat{\mathbf{a}}}P_E(\mathbf{a}\rightarrow \hat{\mathbf{a}})d(\mathbf{a},\hat{\mathbf{a}}),
\label{ber_ub_eq}
\end{equation}
where $f=Q^MM\mathrm{log}_2Q+P\mathrm{log}_2P$, $d(\mathbf{a},\hat{\mathbf{a}})$ corresponds to the number of differing bits between $\mathbf{a}$ and $\hat{\mathbf{a}}$, and $P_E(\mathbf{a}\rightarrow \hat{\mathbf{a}})$ is
\begin{align}
P_E(\mathbf{a}\rightarrow \hat{\mathbf{a}})\approx \frac{1}{12}\left [\left(\prod\nolimits_{r=1}^R\lambda_r\right)^{\frac{1}{R}}\frac{\gamma}{4L}\right]^{-R}+\notag\nonumber\\
\frac{1}{4}\left[\left(\prod\nolimits_{r=1}^R\lambda_r\right)^{\frac{1}{R}}\frac{\gamma}{3L}\right]^{-R}.
\end{align}
The diversity order for DFT-s-OFDM-CM is expressed as
\begin{equation}
G_D=\min_{\mathbf{a},\hat{\mathbf{a}},\mathbf{a}\neq\hat{\mathbf{a}}}\rm{rank}(\mathbf{\Theta}(\mathbf{a},\hat{\mathbf{a}})).
\label{eq_do}
\end{equation}

Considering 3-path equal-gain channel ($L=3$), the maximum Doppler frequency is set to $f_{\rm{max}}=2\,\rm{KHz}$, corresponding to velocity of $v=500\,\rm{km/h}$ and carrier frequency of $f_{\rm{c}}=4\,\rm{GHz}$. The Doppler shift of each user is randomly selected from $-f_{\rm{max}}$ to $f_{\rm{max}}$ for each simulation. The subcarrier spacing is set to $\Delta f=15\,\rm{KHz}$. The sizes of IFFT and DFT for DFT-s-OFDM-CM are set to $N=8$ and $M=2$. Binary phase shift keying (BPSK) modulation with $Q=2$ is considered. The optimized chirp modulation order is found to be $P^{\star}=4$ and $P^{\star}=2$ for $U=1$ and $U=4$, respectively. The chirp modulation order is thus chosen as $P=2$ for both $U=1$ and $U=4$. The simulated BERs of proposed DFT-s-OFDM-CM are shown in Fig. \ref{ber_ub}, where the derived BER upper bound for $U=1$ in \eqref{ber_ub_eq} is served as a benchmark. The simulated BERs for $U=1$ and $U=4$ are observed to approach the derived upper bound particularly at medium to high $E_{\mathrm{b}}/N_0$. Meanwhile, the diversity order calculated using \eqref{eq_do} corresponds to the number of channel paths $L=3$, which aligns with the BER slope in Fig. \ref{ber_ub}. Hence, the proposed DFT-s-OFDM-CM, as a variant of chirped DFT-s-OFDM \cite{10897935}, is able to leverage full frequency diversity in the given simulation scenarios. Due to space limitations, a detailed investigation of the conditions required to achieve full frequency diversity across a broader range of scenarios is left for future work.

\begin{figure}[htbp]
	\centering
	\includegraphics[width=8cm]{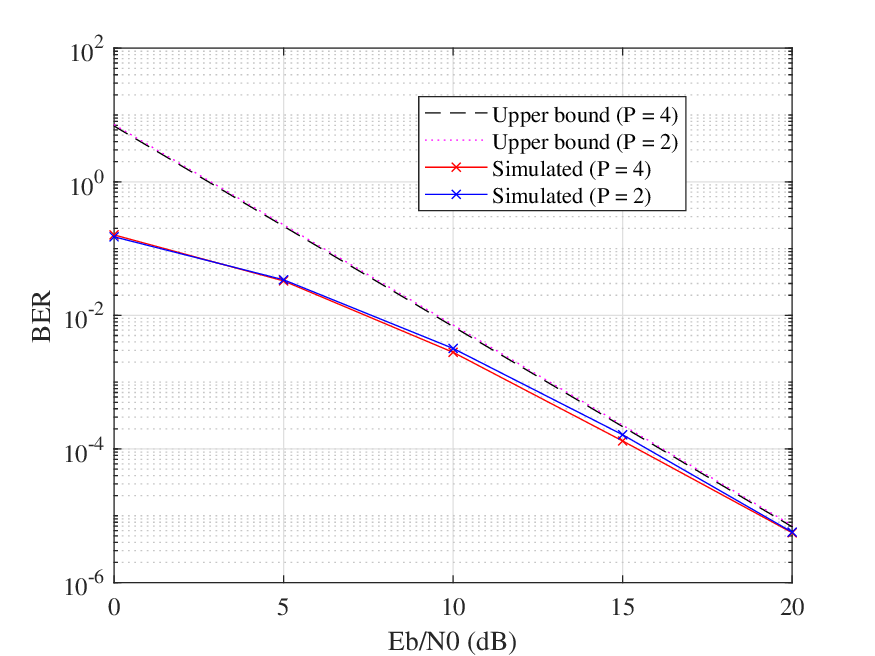}
	\caption{Impact of chirp modulation order $P$ on the derived BER upper bounds and simulated BERs of proposed DFT-s-OFDM-CM, with $N=8$, $M=2$, $Q=2$, and $U=1$.}
\label{ber_ub_p}
\end{figure}

Considering $U=1$, Fig. \ref{ber_ub_p} shows the derived BER upper bounds and simulated BERs for $P=2$ and $P=P^{\star}=4$. When increasing the chirp modulation order $P$ from $2$ to $4$, the simulated BERs and derived BER upper bounds nearly maintain the same. This implies that increasing the chirp modulation order to its optimized value could enhance spectral efficiency with a negligible impact to the BERs.

\section{Simulation Results}
The performance of proposed DFT-s-OFDM-CM is evaluated and compared with DFT-s-OFDM \cite{7744818} and chirped DFT-s-OFDM \cite{10897935} using ML equalizer. The simulation setting similar to Section III is considered. The 3-path equal-gain channel is adopted, with $f_{\rm{max}}=2\,\rm{KHz}$, $v=500\,\rm{km/h}$, and $f_{\rm{c}}=4\,\rm{GHz}$. The Doppler shift of each user is randomly chosen between $-f_{\rm{max}}$ and $f_{\rm{max}}$ for each Monte Carlo simulation.

\begin{figure}[htbp]
	\centering
	\includegraphics[width=8cm]{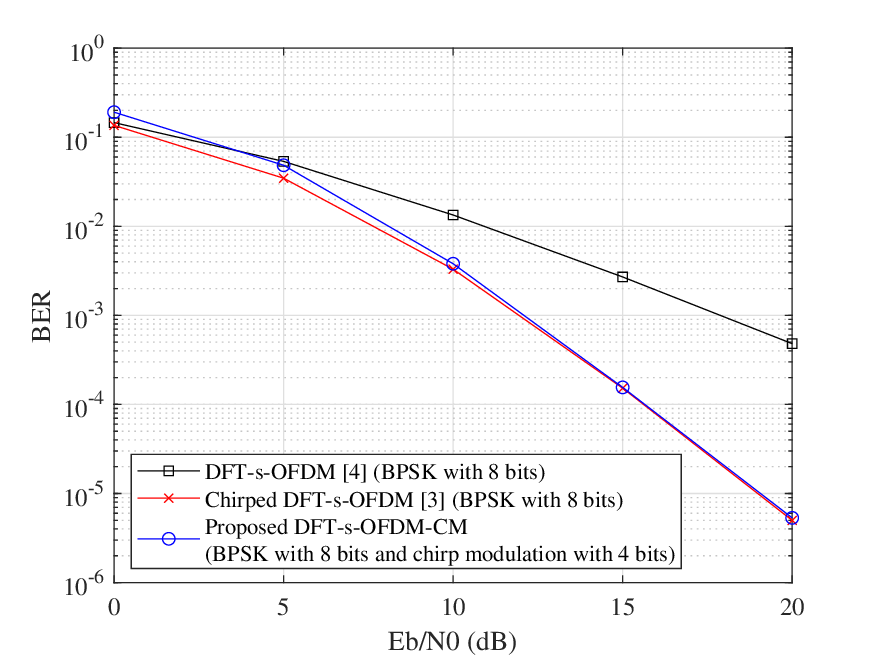}
	\caption{BERs of DFT-s-OFDM \cite{7744818}, chirped DFT-s-OFDM \cite{10897935}, and proposed DFT-s-OFDM-CM, with $N=8$, $M=2$, $P=2$, $U=4$, and $Q=2$.}
\label{ber_diff_wav}
\end{figure}

Fig. \ref{ber_diff_wav} shows the BERs of DFT-s-OFDM \cite{7744818}, chirped DFT-s-OFDM \cite{10897935} and proposed DFT-s-OFDM-CM, with $N=8$, $M=2$, and $U=4$. Note that (chirped) DFT-s-OFDM employs BPSK ($Q=2$) modulation and transmits $8$ bits in total. In addition to the aforementioned $8$ bits, the proposed DFT-s-OFDM-CM transmits extra $4$ bits using chirp modulation with $P=2$, and thus, its total number of bits is $12$. As a result, the spectral efficiency of proposed DFT-s-OFDM-CM is $50\%$ higher than that of (chirped) DFT-s-OFDM. The three waveforms are compared under the same bit to noise ratio ($E_{\mathrm{b}}/N_0$) in Fig. \ref{ber_diff_wav}. The proposed DFT-s-OFDM-CM exhibits the similar BER to that of chirped DFT-s-OFDM \cite{10897935} and both outperforms DFT-s-OFDM \cite{7744818}. Thus, by introducing chirping modulation to DFT-s-OFDM, the proposed DFT-s-OFDM-CM could maintain the superior performance of chirped DFT-s-OFDM \cite{10897935}, while achieving higher spectral efficiency.

\begin{figure}[htbp]
	\centering
	\includegraphics[width=8cm]{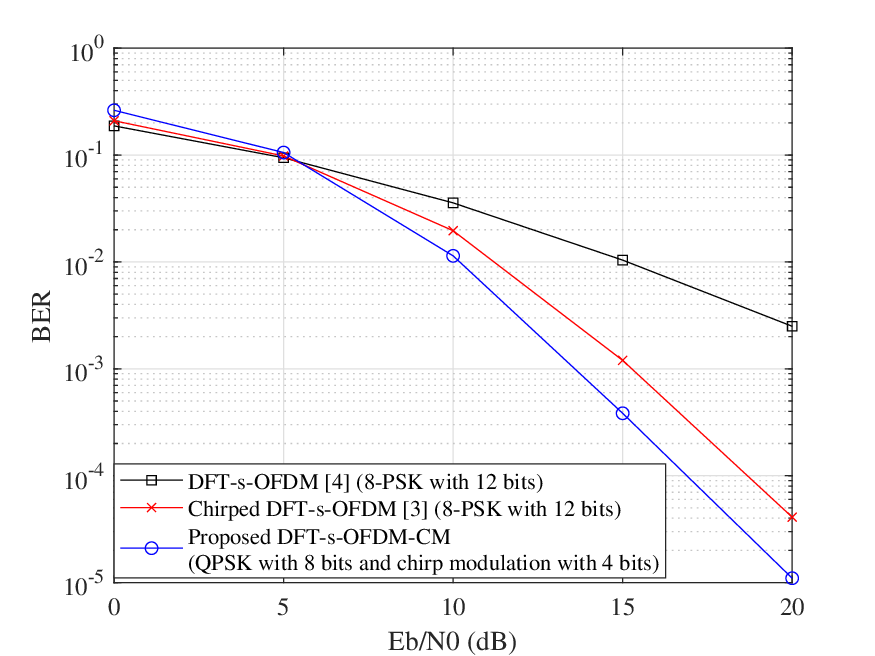}
	\caption{BERs of (chirped) DFT-s-OFDM and proposed DFT-s-OFDM-CM under the same spectral efficiency, with $N=4$, $M=1$, and $U=4$.}
\label{ber_diff_wav_same_se}
\end{figure}

Fig. \ref{ber_diff_wav_same_se} presents the BERs of DFT-s-OFDM \cite{7744818}, chirped DFT-s-OFDM \cite{10897935} and proposed DFT-s-OFDM-CM, with $N=4$, $M=1$, and $U=4$. (Chirped) DFT-s-OFDM adopts 8-PSK modulation with $Q=8$, and thus, the total number of information bits is $UM\log_2 Q=4\times1\times \log_2 8=12$. The proposed DFT-s-OFDM-CM considers quadrature PSK (QPSK) modulation with $Q=4$ and chirp modulation with $P=2$, yielding the total number of information bits $UM\log_2 Q+U\log_2 P=4\times 1\times\log_2 4+4\times1\times\log_2 2=8+4=12$. As a result, all waveforms transmit $12$ bits in total, maintaining the same spectral efficiency. By dividing the number of bits into two streams, lower-order constellation modulation $(Q=4)$ could be enabled for the proposed DFT-s-OFDM-CM, unlike $Q=8$ for (chirped) DFT-s-OFDM. Hence, the proposed waveform could exhibit higher resilience to noise. It can be seen that the proposed DFT-s-OFDM-CM can achieve approximately $2\,\rm{dB}$ and $8\,\rm{dB}$ gain over the chirped DFT-s-OFDM \cite{10897935} and DFT-s-OFDM \cite{7744818}, respectively.

The proposed idea with chirp modulation can be extended to AFDM by encoding the information bits on the starting frequency of its first chirp signal, resulting in the new waveform \mbox{AFDM} with chirp modulation (AFDM-CM). Considering uplink multiple access communications, the transmit time-domain AFDM-CM signal of user $u$ is expressed as
 \begin{equation}
   \mathbf{s}_{\mathrm{afdm-cm},u}=\mathbf{C}_{1,u}\mathbf{F}_N^H\mathbf{C}_2\mathbf{P}_u\mathbf{x}_u,
\end{equation}
where $\mathbf{C}_{1,u}$ and $\mathbf{C}_2$ are $N\times N$ diagonal matrices corresponding to the first and second chirp signal, and $\mathbf{x}_u$ the AFDM signal vector of user $u$ of length $M$. The chirp bits are encoded on $\mathbf{C}_{1,u}$ in a similar manner to that of proposed DFT-s-OFDM-CM as shown in Eq. \eqref{c_u_n_circ}. The proposed AFDM-CM differs from the existing AFDM-related studies \cite{10975107,10943004} that the information bits are encoded on the first chirp signal rather than the second chirp signal of AFDM. Additionally, it circularly shifts the first chirp signal instead of permuting the second chirp signal as in AFDM-CPIM \cite{10943004}.

\begin{figure}[htbp]
	\centering
	\includegraphics[width=8cm]{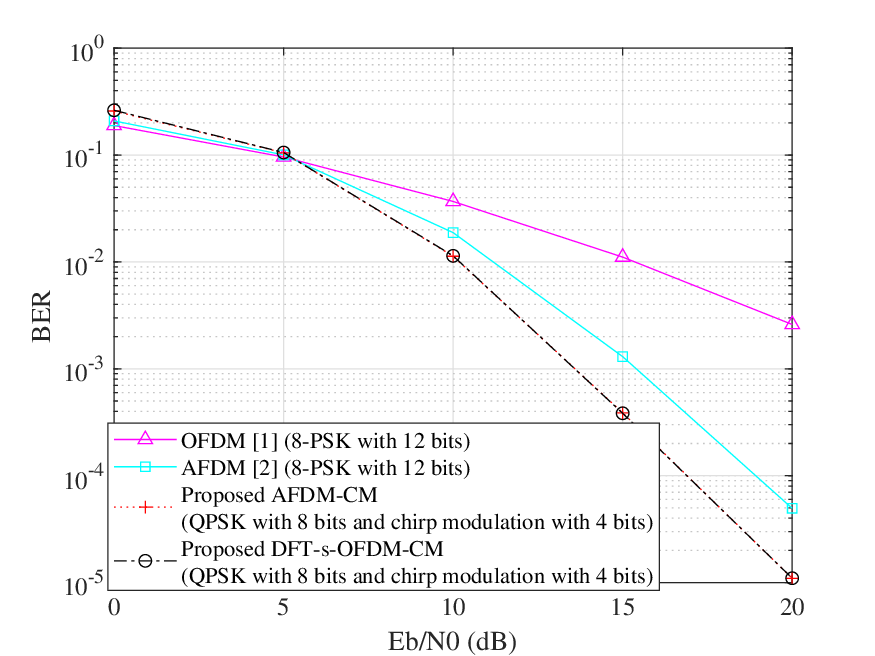}
	\caption{BERs of AFDM \cite{afdm_twc}, OFDM \cite{9468975}, proposed AFDM-CM, and proposed DFT-s-OFDM-CM, with $N=4$, $M=1$, and $U=4$.}
\label{ber_afdm}
\end{figure}

Considering the same simulation setting to that of Fig.~\ref{ber_diff_wav_same_se}, Fig. \ref{ber_afdm} shows the BER of proposed AFDM-CM in comparison to that of proposed DFT-s-OFDM-CM, AFDM \cite{afdm_twc} and OFDM~\cite{9468975} based systems, with $N=4$, $M=1$, and $U=4$. AFDM-CM and DFT-s-OFDM-CM employ QPSK modulation with $Q=4$ and chirp modulation with $P=2$, while AFDM and OFDM utilize 8-PSK modulation with $Q=8$. These four waveforms transmit $12$ bits and maintain the same spectral efficiency. Similarly to Fig. \ref{ber_diff_wav_same_se}, the proposed AFDM-CM enables a smaller constellation modulation order, making it less susceptible to noise. It achieves approximately $2\,\rm{dB}$ and $8\,\rm{dB}$ gain over that of AFDM \cite{afdm_twc} and OFDM~\cite{9468975}.

Note that DFT-s-OFDM-CM and AFDM-CM exhibit the similar BERs in Fig. \ref{ber_afdm}. However, similar to Fig. 4a in \cite{10897935}, the variant of chirped DFT-s-OFDM would achieve lower PAPR than the variant of AFDM. The PAPR of AFDM-CM would depend on the number of subcarriers, while that of DFT-s-OFDM-CM is determined by the $Q$-ary constellation symbol. According to Fig. 6b in \cite{10897935}, the modulation waveform with lower PAPR, \emph{i.e.}, DFT-s-OFDM-CM, would exhibit higher resilience to clipping than AFDM-CM.


\section{Conclusion}
In this paper, DFT-s-OFDM-CM and AFDM-CM are proposed for the next generation of communications by embedding information bits on both the $Q$-ary constellation symbols and the starting frequency of chirp signal. The proposed DFT-s-OFDM-CM could enhance spectral efficiency, while achieving the similar BER to that of chirped DFT-s-OFDM~\cite{10897935}. It is also able to maintain low PAPR and fully exploit frequency diversity. When maintaining the same spectral efficiency, the two proposed waveforms with the splitting of information bits into two streams exhibit higher resilience to noise and thus lower BER than AFDM~\cite{afdm_twc} and chirped DFT-s-OFDM~\cite{10897935}.

\bibliographystyle{IEEEtran}
{\small
\bibliography{references}}
\addcontentsline{toc}{section}{References}

\IEEEpeerreviewmaketitle

\end{document}